\documentclass{webofc}
\usepackage[varg]{txfonts}   % Web of Conferences font
\usepackage{listings}
\usepackage{hyperref}
\begin{document}
\title{Pulsating stars in the VMC survey\thanks{Based on observations made with VISTA at ESO under programme ID 179.B-2003.}}
%
% subtitle is optional

\author{\firstname{Maria-Rosa L.} \lastname{Cioni}\inst{1,2}\fnsep\thanks{\href{mailto:mcioni@aip.de}{\tt mcioni@aip.de}} \and
        \firstname{Vincenzo} \lastname{Ripepi}\inst{3} \and 
        \firstname{Gisella} \lastname{Clementini}\inst{4} \and
        \firstname{Martin A.T.} \lastname{Groenewegen}\inst{5} \and
        \firstname{Maria I.} \lastname{Moretti}\inst{3} \and
        \firstname{Tatiana} \lastname{Muraveva}\inst{4}  \and
        \firstname{Smitha} \lastname{Subramanian}\inst{6}       
}

\institute{Leibniz-Instit\"{u}t f\"{u}r Astrophysik Potsdam, An der Sternwarte 16. 14482 Potsdam. Germany
\and
University of Hertfordshire, PAM, College Lane, Hatfield AL10 9AB, United Kingdom
\and
INAF-Osservatorio Astronomico di Capodimonte, Salita Moiariello 16, 80131 Napoli, Italy
\and
INAF-Osservatorio Astronomico di Bologna, via Gobetti 93/3, 40129 Bologna, Italy
\and
Koninklijke Sterrenwacht van Belgi\"{e}, Ringlaan 3, 1180 Brussels, Belgium
\and
Kavli Institute for Astronomy and Astrophysics, Peking University, Yi He Yuan Lu 5, Hai Dian District, Beijing 100871, China
          }

\abstract{%
  The VISTA survey of the Magellanic Clouds system (VMC) began observations in 2009 and since then, it has collected multi-epoch data at $K_\mathrm{s}$ and in addition multi-band data in $Y$ and $J$ for a wide range of stellar populations across the Magellanic system. Among them are pulsating variable stars: Cepheids, RR Lyrae, and asymptotic giant branch stars that represent useful tracers of the host system geometry.
}
\maketitle

\section{Introduction}\label{sec:intro}

The Magellanic Clouds are prominent sky markers in the southern hemisphere since the time of the first circumnavigation of the globe by Fern\~{a}o de Magalh\~{a}es in 1519-1522 (\cite{pigafetta1994}). They are the most luminous and largest dwarf satellite galaxies of the Milky Way. They are metal poor, nearby, and are interacting with each other and with the Milky Way. Signatures of these interactions are associated to a Bridge and Stream; both features are predominately traced by neutral hydrogen that was stripped from the galaxies as a result of dynamical interactions. Most likely they were accreted only recently (about $2$ Gyr ago) and arrived under the influence of the Milky Way potential together with their own satellites (\cite{koposov2015}).

\section{The VMC survey}\label{sec:sec-1}

The near-infrared $YJK_\mathrm{s}$ VISTA ESO public survey of the Magellanic Clouds system (VMC) began in 2009 and is expected to be completed in 2018. This is a photometric survey in three filters, $Y$, $J$, and $K_\mathrm{s}$ performed at the VISTA telescope using the VIRCAM camera. The latter provides a spatial resolution of $0.34^{\prime\prime}$ per pixel and a field-of-view of $1.65$ deg$^2$ sampled by $16$ Rayethon detectors. Details about the telescope and instrument are given in \cite{vista}. The VMC survey covers and area of approximately $170$ deg$^2$ of the Magellanic system and includes stars as faint as $\sim 22$ mag ($5\sigma$ Vega) at each filter. Stars brighter than $\sim10$ mag at $K_\mathrm{s}$ are usually saturated. The survey provides three independent epochs at the $Y$ and $J$ filters and at least $12$ independent epochs at the $K_\mathrm{s}$ filter. Details about the survey strategy are given in \cite{cioni2011}.

\begin{figure*}
\centering
\includegraphics[width=12cm,clip]{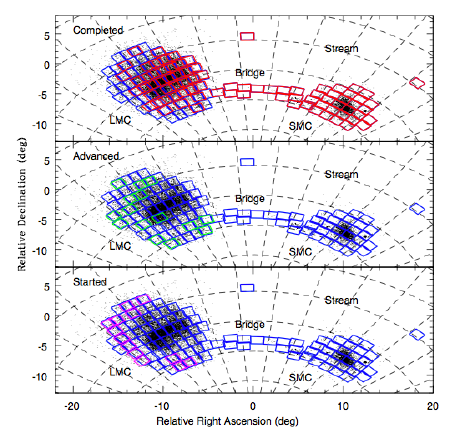}
\caption{Progress of VMC observations. Rectangles identify individual VMC tiles as follows: all $110$ survey tiles (blue), tiles fully observed at all three filters and epochs (red), tiles that are in an advanced stage of observation, namely tiles for which most of the $YJ$ observations and at least $6$ epochs at $K_\mathrm{s}$ have been obtained (green), and tiles that have only a few observations acquired (magenta). The image is centred at ($\alpha=03$:$24$:$00$, $\delta=-69$:$00$:$00$) and dashed lines indicate steps of $40$ min and $5$ deg in RA and DEC, respectively.} 
\label{vmc}       
\end{figure*}

The main scientific goals of the VMC survey are to derive the spatially resolved star formation history across the Magellanic system and to measure its three-dimensional geometry. The first goal defines the depth of the survey because it is necessary to observe stars as faint as below the old main-sequence turnoff to derive age and metallicity from the interpretation of colour-magnitude diagrams using theoretical models. The second goal defines the monitoring strategy of the survey in order to derive distances from variable stars (Cepheids and RR Lyrae stars); red clump giant stars are also used to measure distances. The VMC survey also addresses a wide range of legacy science: from the study of individual objects in the Magellanic Clouds and in the Milky Way, star formation, stellar clusters, the dynamics of the Magellanic system, to specific investigations of background galaxies and quasars.

The effective area covered in one observation (a pawprint) is not contiguous because of large gaps among the detectors. To cover homogeneously the field-of-view (a tile) it is necessary to fill these gaps with a six-point mosaic. In this way pixels in the central region of $1.5$ deg$^2$ in size are observed $2-6$ times while in two external regions (tile wings) they are observed only once, within a given tile. The VMC survey covers the Magellanic system with 110 overlapping tiles (Fig. \ref{vmc}). The spacing between the $K_\mathrm{s}$ epochs follows a cadence with a minimum spacing of $0$, $1$, $3$, $7$, and $17$ days for each subsequent epoch, while there is no requirement on the spacing of the $Y$ and $J$ band epochs; these can all be obtained during the same night. To observe one VMC tile, in a given filter, it takes about $1$ hour. The overall time to complete the VMC survey is of $\sim 2000$ hours or $\sim 250$ nights. Observations for the VMC survey are obtained in service mode which guarantees homogeneity in sky quality (clouds, airmass, and seeing).

The VMC data are first processed at the Cambridge Astronomy Survey Unit using the VISTA Data Flow System pipeline (VDFS; \cite{irwin2004}). Pawprint and tile images are produced and the corresponding source catalogues per filter and epoch are created. Next, deep images resulting from stacking individual epoch observations, with their associated catalogues are created at the Wide-Field Astronomy Unit. Individual epochs are also linked with each other. The VMC data are ingested in the VISTA science archive (VSA; \cite{cross2012}) which is a point of access for both the VMC team and the astronomical community. At yearly intervals, some VMC data are made publicly available and they can be accessed at the VSA or at the ESO archive. The VMC team creates additional data products, such as catalogues with point spread function (PSF) photometry and catalogues of variable stars, which are also released to the community via the two points of entry.

The VMC public web page is at http://star.herts.ac.uk/$\sim$mcioni/vmc where it is possible to view the deep three-colour images with a user friendly zoom-in facility that allows us to appreciate the richness and quality of the data. As of $10^\mathrm{th}$ March 2017 the VMC observations are $88\%$ complete. In particular, $83$ tiles are fully observed in each filter and epoch, the SMC and the Bridge regions are also fully observed, and the LMC is $\sim 80\%$ complete (Fig. \ref{vmc}). The fourth public data release of the VMC data comprises of $19$ tiles ($10$ in the LMC, $5$ in the SMC, $2$ in the Bridge, and $2$ in the Stream). For these tiles deep images in each filter with their associated pipeline and PSF catalogues, single tile images per filter with associated source lists, and individual pawprint images also with their associated catalogues are made available, as well as catalogues of some variable stars.

\section{VMC results on pulsating stars}\label{sec:sec-2}

\subsection{Cepheids}\label{sec:cepheids}

Cepheids are hydrogen-shell burning stars crossing the instability strip. They are often classified as: classical or type I Cepheids that refer to $4-20$ M$_\odot$ stars a few $100$ Myr old, anomalous Cepheids that refer to $1.3-2.1$ M$_\odot$ stars a few Gyr old and metal poor and type II Cepheids that refer to $0.5$ M$_\odot$ stars $9-10$ Gyr old. Cepheids are important standard candles via period-luminosity, period-luminosity-colour, and Wesenheit relations. In the near-infrared chemical effects (from He and Fe) on these relations are reduced, reddening is low, light-curves are sinusoidal, and the relations are linear with a small dispersion. These properties represent clear advantages in using these stars and near-infrared pass-bands to measure distances.

\begin{figure}
\sidecaption
\centering
\includegraphics[width=7cm,clip]{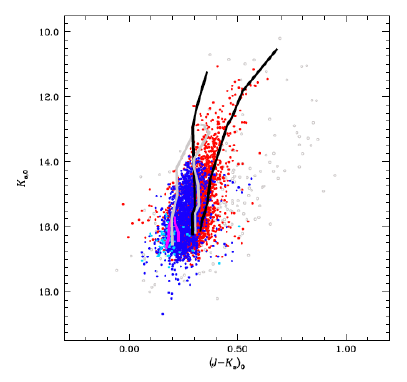}
\caption{Color-magnitude diagram of classical Cepheids in the SMC. Theoretical instability strips for fundamental (black lines; red points), first overtone (grey lines; blue points), and second overtone (magenta lines; cyan points) pulsators are indicated. See Ripepi et al. (\cite{ripepi2016}) for details.} 
\label{strip}       
\end{figure}

\begin{figure}
\centering
\includegraphics[width=7cm,clip]{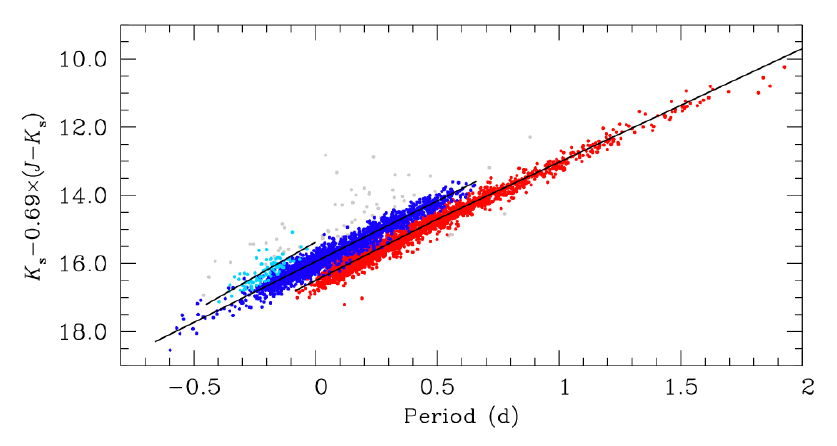}
\includegraphics[width=7cm,clip]{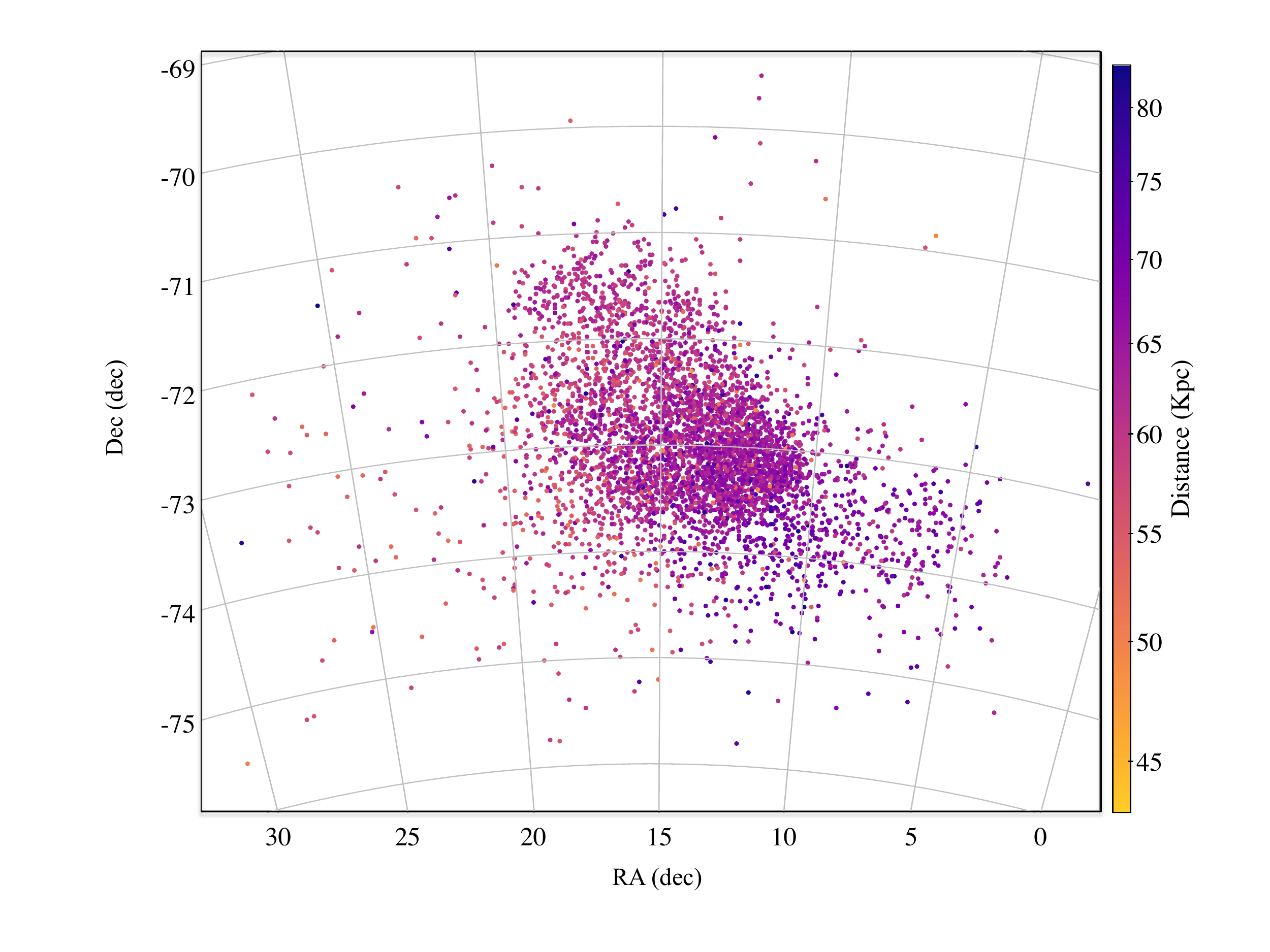}
\caption{(Left) Period-Wesenheit relation for classical Cepheids pulsating in the fundamental (red), first overtone (blue), and second overtone (cyan) modes. The scatter represents the intrinsic depth of the SMC. (Right) Spatial distribution of classical Cepheids across the SMC; points are colour coded by distance as indicated in the legend. Details are given in Ripepi et al. (in preparation).} 
\label{smcpl}       
\end{figure}

In the ($J-K_\mathrm{s}$, $K_\mathrm{s}$) diagram (Fig. \ref{strip}), the instability strip for the combined modes of pulsation (fundamental, first and second overtone) spans about $0.5$ mag in colour and $8$ mag in $K_\mathrm{s}$ band. Using VMC multi-epoch data up to eight different light-curve templates were built and subsequently used to measure accurate mean magnitudes for the Cepheids (\cite{ripepi2016}). Combining these magnitudes with the period obtained from the Optical Gravitational Lensing Experiment (OGLE) project we constructed period-luminosity relations and alike with a small scatter (Fig. \ref{smcpl} - left). Considering that the depth of the Magellanic Clouds is about $0.2$ mag for the LMC and $0.7$ mag for the SMC, we are able to use these relations to quantify distance effects throughout the galaxies. Preliminary maps of the spatial distribution of classical Cepheids in the different combination of cartesian coordinates indicate that the de-projected SMC resembles the LMC galaxy with an elongated bar and a region above it that can be associated to a spiral arm. We confirm a complex structure for the galaxy and a large depth along the line-of-sight ($\sim 20$ kpc) where stars to the northern east region are closer to us, and to the LMC, than stars in the southern west region (Fig. \ref{smcpl} - right); see also (\cite{dobra2016} and \cite{rubele2015}).

\begin{figure*}
\centering
\includegraphics[width=4cm,clip]{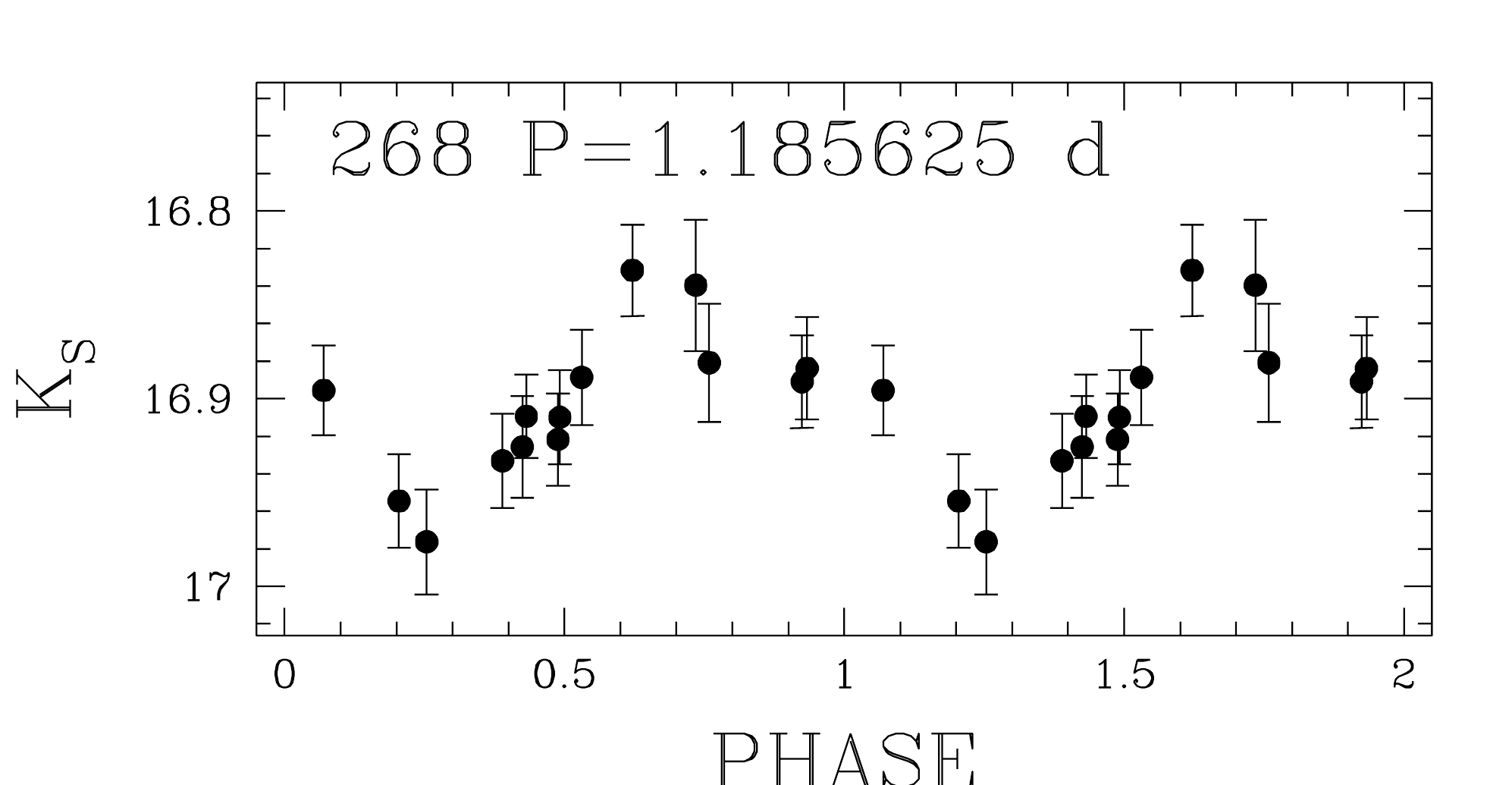}
\includegraphics[width=4cm,clip]{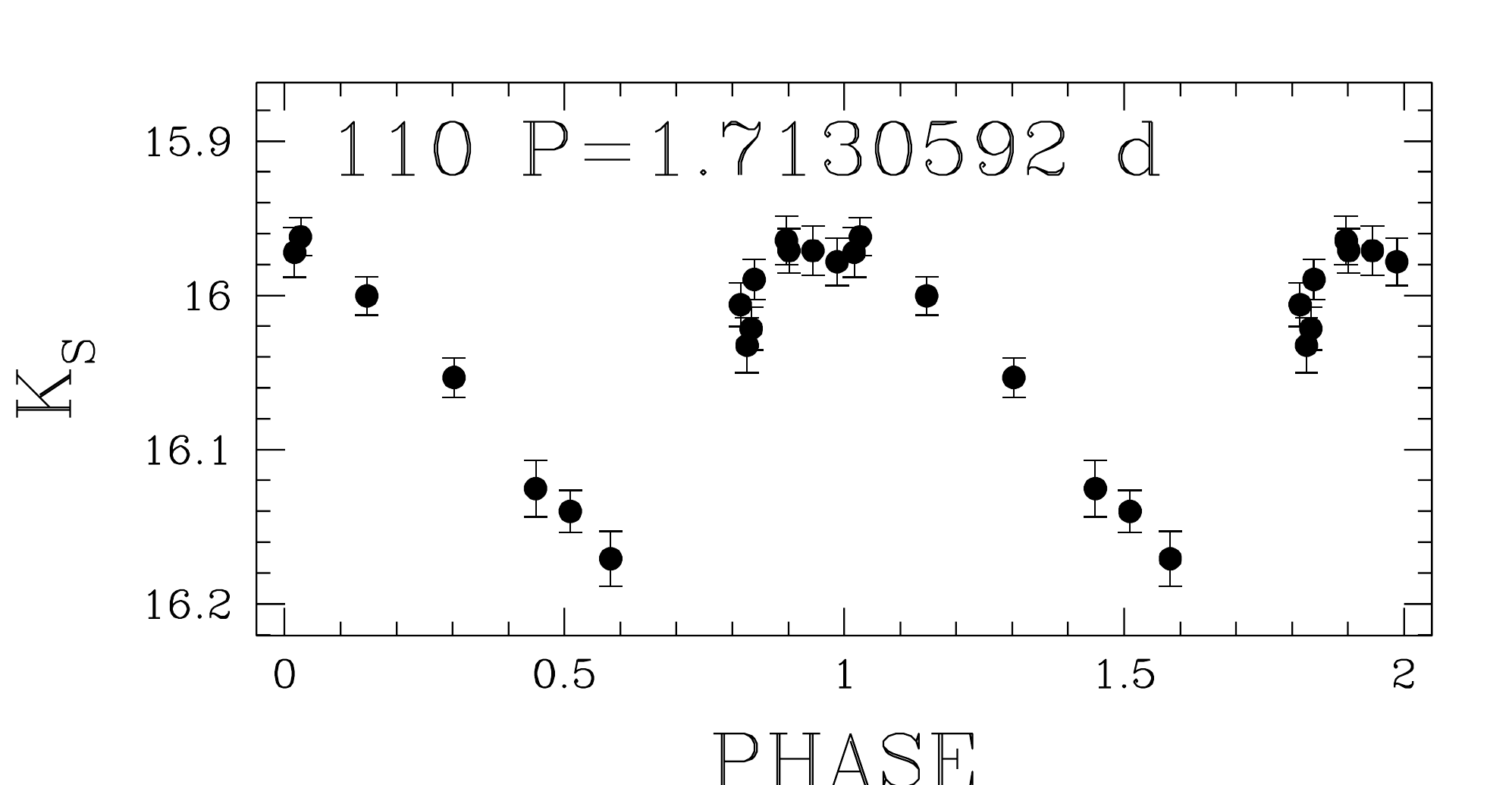}
\includegraphics[width=4cm,clip]{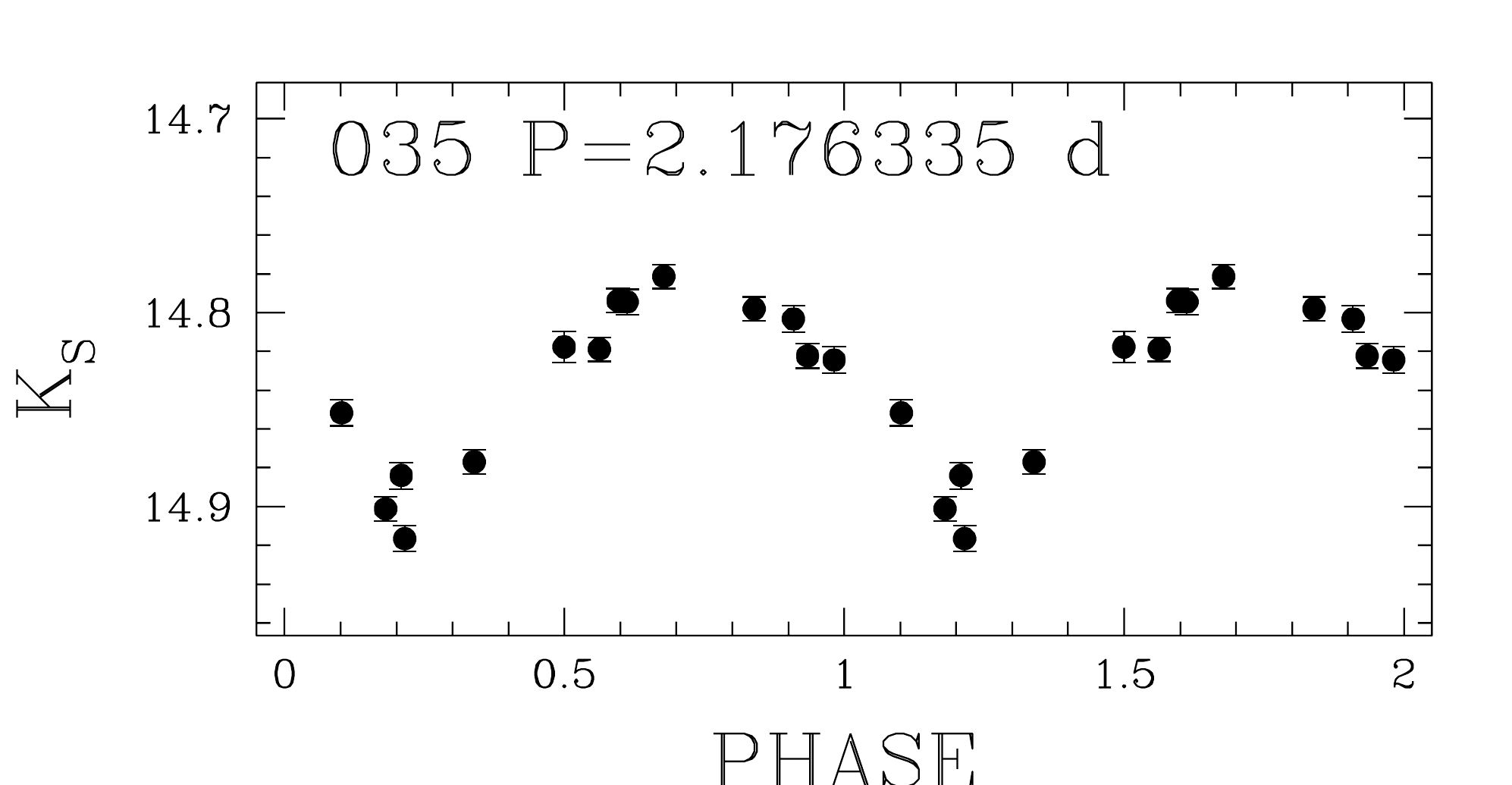} 
\includegraphics[width=4cm,clip]{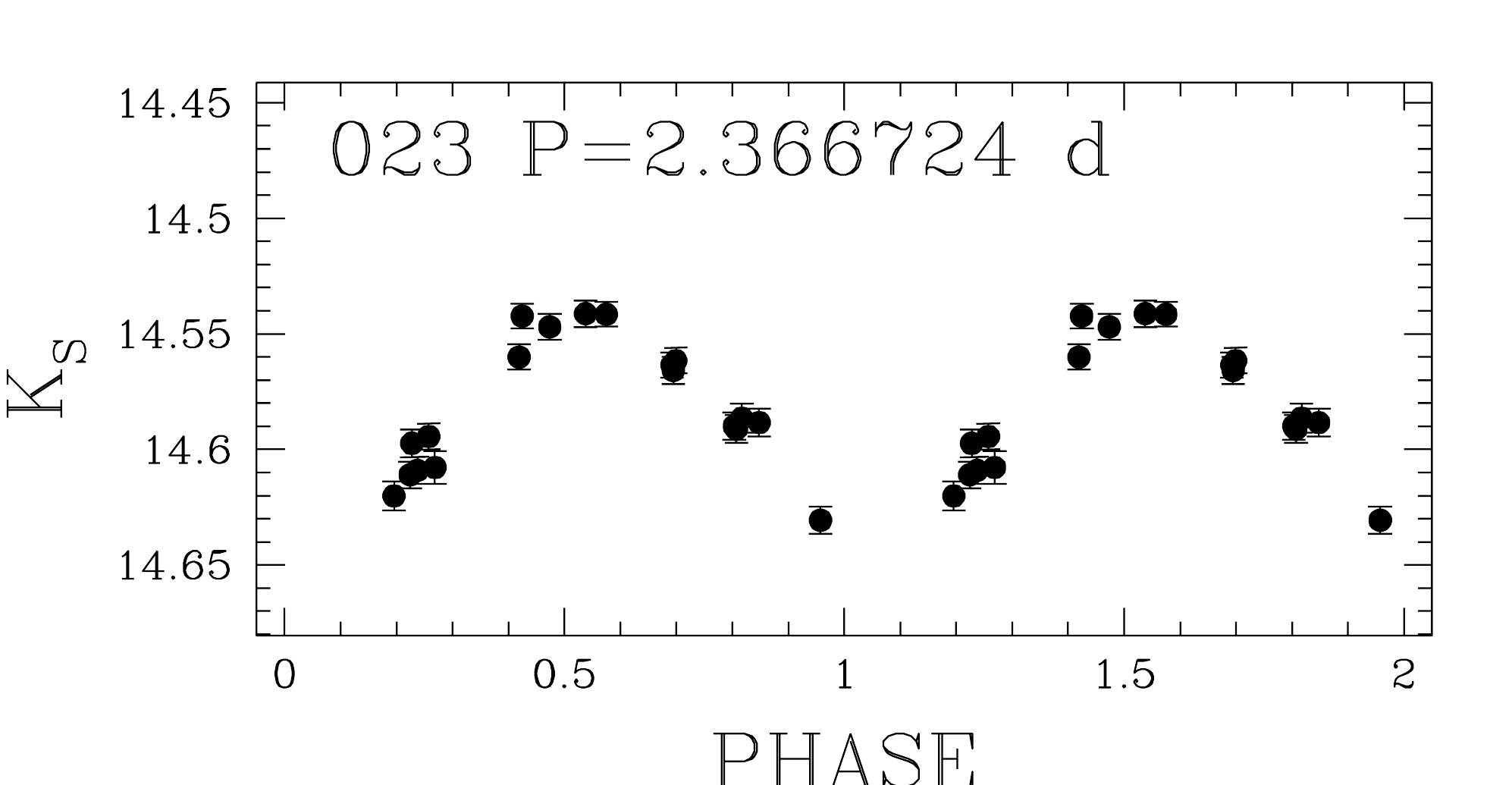}
\includegraphics[width=4cm,clip]{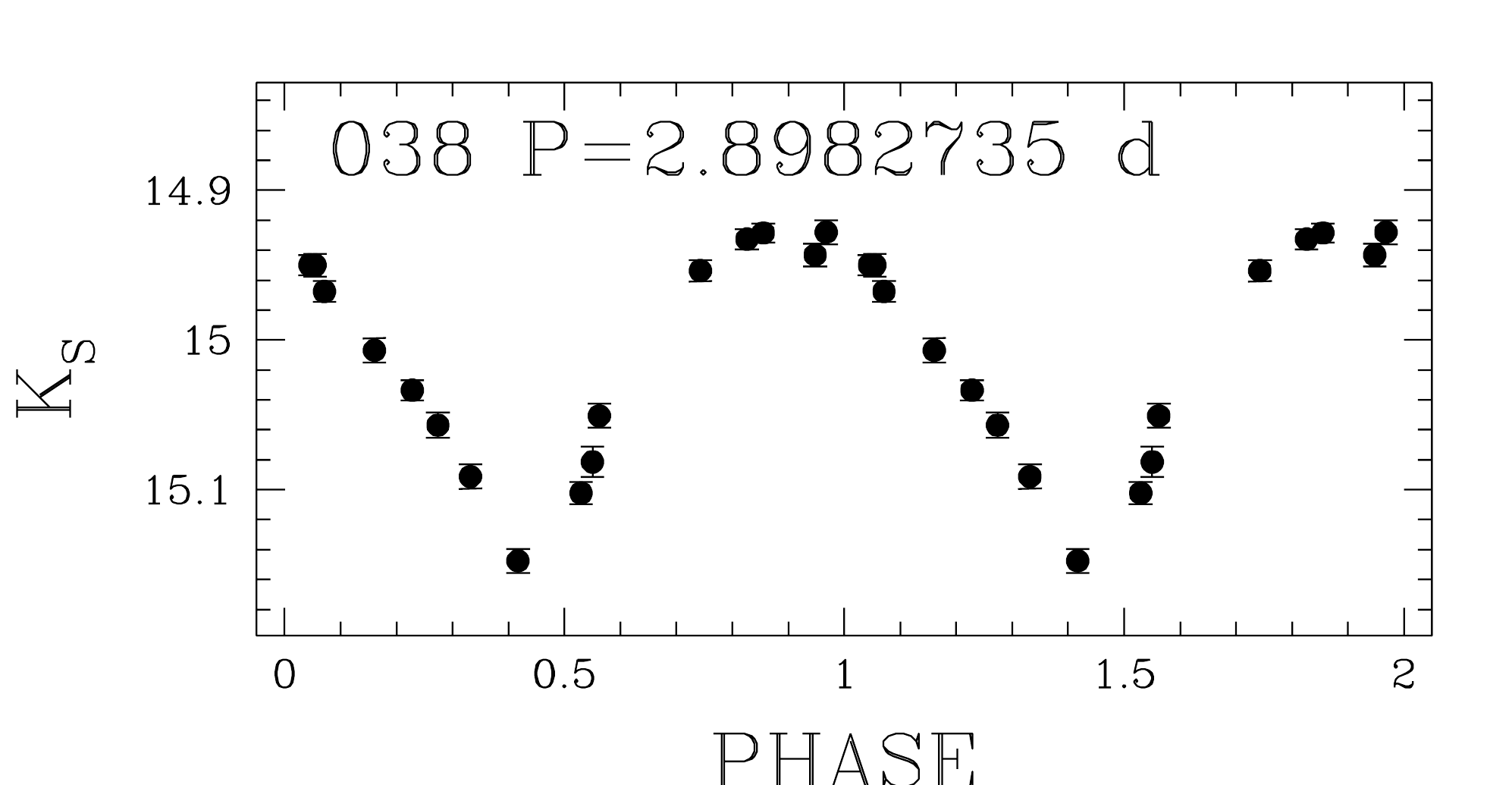}
\includegraphics[width=4cm,clip]{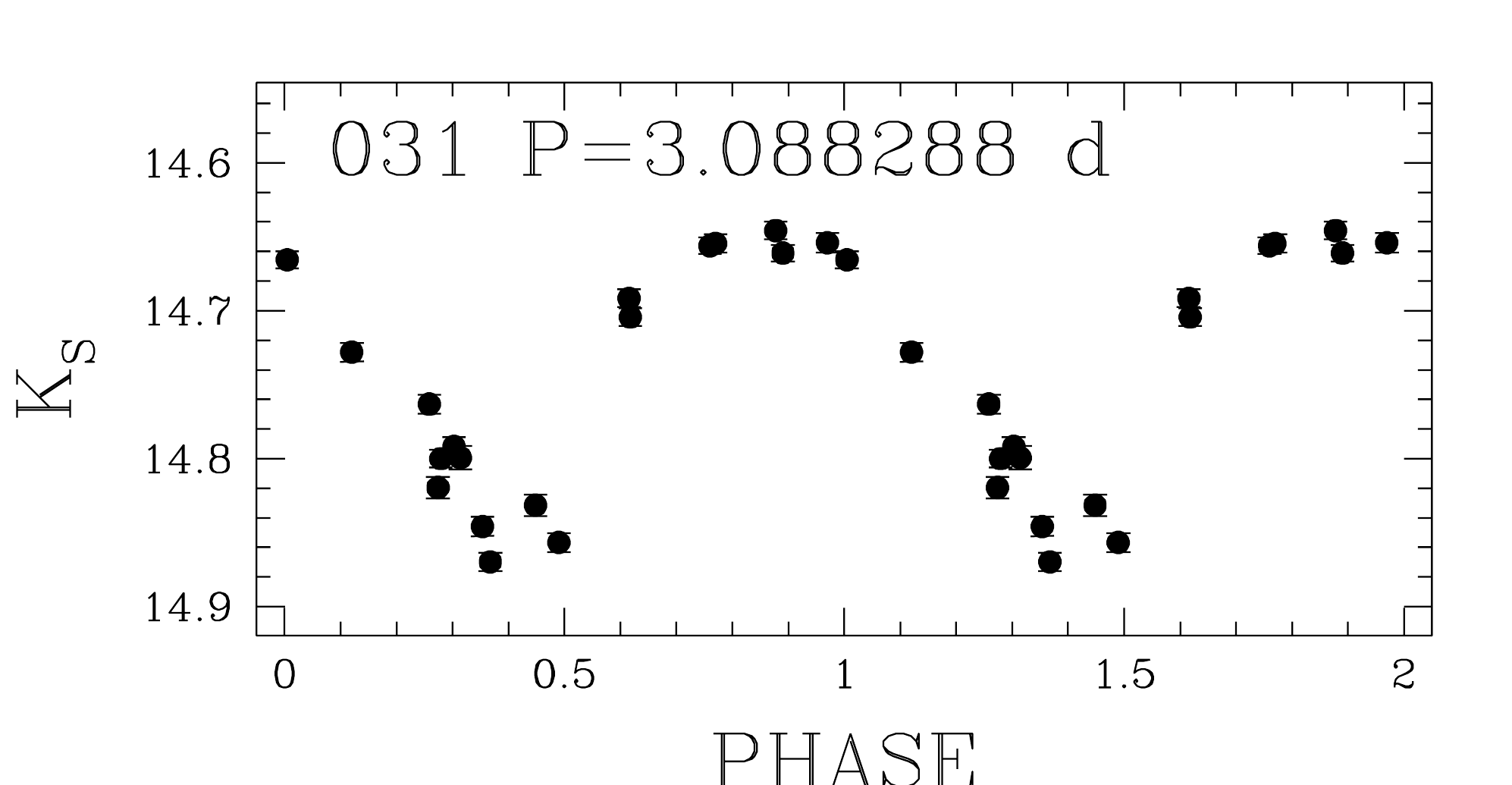} 
\caption{$K_\mathrm{s}$-band curves for six candidate classical cepheids newly discovered by the VMC survey. Reference numbers and the periods of pulsation are indicated in each panel. Further details are given in \cite{moretti2016}.} 
\label{cepheids}       
\end{figure*}

Based on the distribution of known classical Cepheids in the VMC colour-magnitude and colour-colour diagrams, on the period-luminosity relation, and as on the shape of their $K_\mathrm{s}$-band light-curve, we have identified about $290$ new candidate classical Cepheids in the external regions of the SMC, using only the VMC near-infrared photometry (\cite{moretti2016}) . A few light-curves are shown in Fig. \ref{cepheids}. The number of SMC Cepheids we have detected is consistent with the predictions of Rubele et al. (\cite{rubele2015}) SFH recovery in these regions of the galaxy. About $13\%$ of them were subsequently confirmed by the OGLE IV project.

VMC $JK_\mathrm{s}$ photometry of nine classical Cepheids in the SMC was used together with optical photometry and radial velocity data to find best fit models that provides distances, masses, luminosities, and radii of the stars (Marconi et al. \cite{marconi2017}; see also this proceeding).

\subsection{RR Lyrae stars}\label{sec:rrlyraestars}

\begin{figure}
\includegraphics[width=7cm,clip]{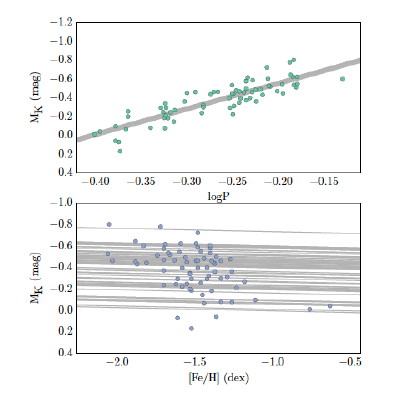}
\includegraphics[width=7cm,clip]{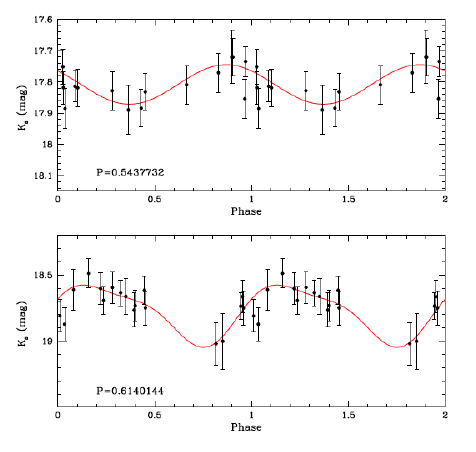}
\caption{(Left) Projections of the  $PK_\mathrm{s}Z$ relation of RR Lyrae stars in the LMC on the $\log({\rm P})-K_\mathrm{s}$ (top) and $K_\mathrm{s}-{\rm [Fe/H]}$ (bottom) planes. Grey lines are lines of equal metallicity (top) or equal period (bottom). (Right) Phased $K_\mathrm{s}$-band light curves of SMC RR Lyrae stars observed by the VMC survey. Periods are from the OGLE III catalogue and measured in days. Red lines show the best fits.}
\label{fig:plLMC}
\end{figure}

RR Lyrae are old ($>10$ Gyr), low mass ($0.6-0.8$ M$_\odot$) stars burning helium in their core.  They populated the horizontal branch. The amplitude of their light variation in the $K_\mathrm{s}$ band is significantly smaller than in the visual and they are fainter than Cepheids. This makes them more difficult to detect and study. Nevertheless, we have analysed a sample of $70$ RR Lyrae stars in the LMC for which $K_\mathrm{s}$-band magnitudes from the VMC survey, spectroscopically determined metallicities (Gratton et al. \cite{gratton2004}) and periods from the OGLE III survey are available (\cite{muraveva2015}).  We computed a period-luminosity-metallicity  ($PK_\mathrm{s}Z$) relation for these stars. Figure \ref{fig:plLMC} shows the projections of the derived $PK_\mathrm{s}Z$ relation on the $\log({\rm P})-K_\mathrm{s}$ and $K_\mathrm{s}-{\rm [Fe/H]}$ planes.

\begin{figure}
\includegraphics[width=6cm,clip]{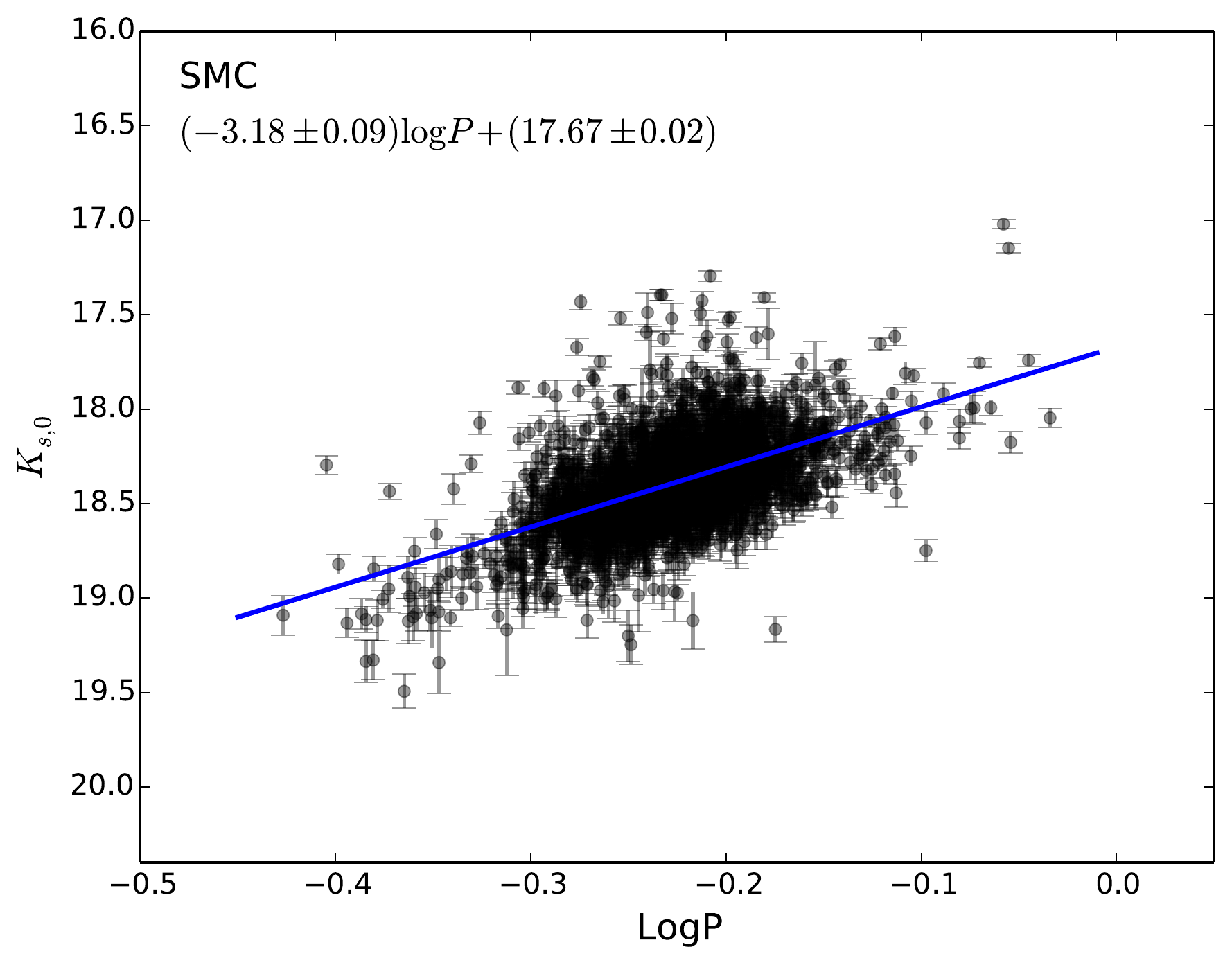}
\includegraphics[width=7.5cm,clip]{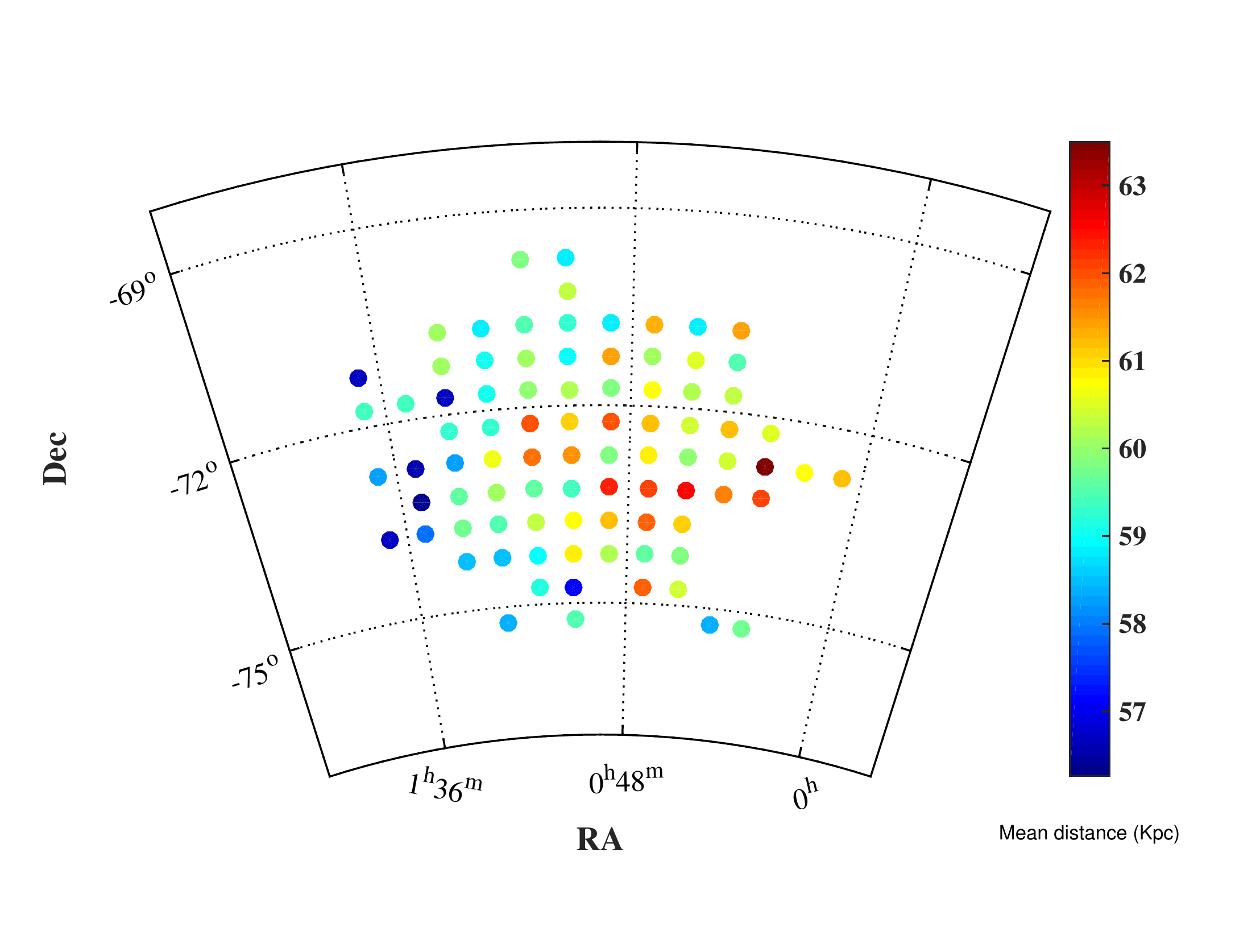}
\caption{(Left) $PK_\mathrm{s,0}$ relation of $2662$ RR Lyrae stars in the SMC. (Right) Two dimensional distribution of mean distance measured in different sub-regions of the SMC. Each point corresponds to an area of $0.6 \times0.5$ deg$^2$. Details are given in Muraveva et al. (in preparation).}
\label{plmap}
\end{figure}

 The derived $PK_\mathrm{s}Z$ relation was used to determine individual distances to $2662$ RR Lyrae stars pulsating in the fundamental mode and distributed across the SMC and, thus, study the structure of this galaxy. Examples of the light-curves of RR Lyrae stars in the SMC are shown in Fig. \ref{plmap}, where the red line represents the best fit.  We have derived mean magnitudes and constructed $PK_\mathrm{s}$  relation for these stars (Fig. \ref{plmap}). The large scatter ($\sim 0.20$ mag) of the relation reflects the extended structure of the SMC along the line-of-sight.  Preliminary results show that RR Lyrae stars in the south-eastern region of the SMC are at closer distance  (Fig. \ref{plmap}) . This is suggestive of tidal interactions. The spatial distribution is almost uniform and the three-dimensional structure is ellipsoidal. The line-of-sight depth varies from $500$ pc to $10$ kpc.

\subsection{Asymptotic giant branch stars}\label{sec:agbstars}

Stars with initial masses up to $8$ M$_\odot$ evolve to the asymptotic giant branch (AGB) phase. In particular, intermediate-age evolved giant stars are classified into M-type or C-type depending on their atmospheric chemical composition. They are also characterised by pulsation with long periods (a few hundred days) and large amplitudes (exceeding one magnitude in some bands). These stars trace the extended structure of the Magellanic Clouds (e.g. \cite{cioni2000}) and have recently been found by Gaia in tidal tails (\cite{belokurov2016}). The VMC survey provides a relatively short sampling of their $K_\mathrm{s}$-band curve, but combined with literature observations from 2MASS and DENIS they allow us to derive the amplitude and period of variation with greater accuracy than by using only fewer and sparser measurements. Ultimately, an accurate measurement of the mean $K_\mathrm{s}$ magnitude of AGB stars will provide better period-magnitude relations that, similarly to those obtained for Cepheids and RR Lyrae stars, will allow us to estimate distances, and trace the geometrical distribution of the Magellanic Clouds at intermediate ages.

\begin{figure}
\centering
\sidecaption
\includegraphics[width=9cm,clip]{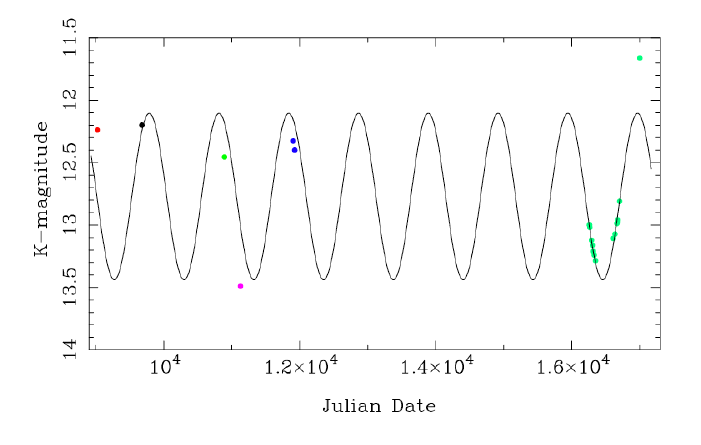}
\caption{$K_\mathrm{s}$-band curve for the AGB star IRAS05506 combining VMC data (cluster of green points) with literature data from other sources (other coloured points). The best fit periodic variation is indicated with a thin grey line. Small shifts are probably due to the different filter curves associated to each survey.}
\label{iras}       
\end{figure}

\section{Conclusions}\label{sec:conclusions}
The VMC observations are at least $80\%$ complete. These unique data produce accurate mean $K_\mathrm{s}$-band magnitudes, for different types of stars, from at least $12$ independent epochs. Within the next years we would like to use this data to go a step further into our understanding of the Magellanic Clouds and of specific stars. The analysis of the structure of the SMC using Cepheids and RR Lyrae stars will soon be finalised and the results will be compared with the recent studies based on the OGLE data. The analysis of the structures of the LMC and of the Bridge will follow and AGB stars will provide a complementary view of the system.

The study of the distribution and properties of stellar populations of different ages will allow us to set constraint on the orbital history of the Magellanic system.
Recent simulations of the two body system, Milky Way and LMC, support the scenario that the LMC is either on a first passage or on a long period orbit (5 Gyr) around the Milky Way (\cite{besla2016}). A comparison with analogous systems in the Illustris simulations support a first approach. In this case the mass of the Milky Way would be limited to $1.5\times10^{12}$ M$_\odot$ and the LMC would have retained most of its cosmological mass. 

To further characterise, dynamically and chemically, the Magellanic system we need to obtain a large set of spectra that in radial velocity match the proper motions obtained at present with VMC and soon with Gaia, that sample substructures and all stellar populations. In addition, iron abundances will permit us to estimate better ages, from the colour-magnitude diagrams and to assess period-luminosity dependencies that may influence the accuracy via which distances are measured. These are the goals of a new large scale project to be undertaken at VISTA with the 4MOST instrument (\cite{dejong2016}). The One Thousands and One Magellanic Fields survey (1001MC; PI Cioni) will obtain both low ($R=5000$) and high $R=20000$) resolution spectra across and area of about $1000$ deg$^2$ for about half a million stars with $r<20$ mag. These data will allow us to measure on average radial velocities with an accuracy $<2$ km/s and iron abundance with an accuracy of $<0.2$ dex. This survey will begin in $2022$ and will last for approximately $5$ years. For details see http://https://www.4most.eu/cms/surveys/galactic/.

\begin{acknowledgement} 
\noindent\vskip 0.2cm
\noindent {\em Acknowledgments}: This project has received funding from the European Research Council (ERC) under European Union's Horizon 2020 research and innovation programme (grant agreement No 682115). We thank the Cambridge Astronomy Survey Unit (CASU) and the Wide Field Astronomy Unit (WFAU) in Edinburgh for providing calibrated data products under the support of the Science and Technology Facility Council (STFC) in the UK. M.-R.L. Cioni acknowledges support from the STFC [grant number ST/M001008/1].
\end{acknowledgement}

% BibTeX or Biber users please use (the style is already called in the class, ensure that the "woc.bst" style is in your local directory)
% \bibliography{name or your bibliography database}

\end{document}